\documentclass[11pt,a4paper]{article}
\usepackage{amsmath}
\usepackage{setspace}
\usepackage{xypic}
\usepackage[all]{xy}
\usepackage{latexsym}
\usepackage{theorem}
\newtheorem{teorema}{Theorem}[section]

\include{amslatex}
\input xy
\newtheorem{proposicion}[teorema]{Proposition}

 \textheight
22 cm \textwidth 14 cm \oddsidemargin 7.5 mm \topmargin 0 mm
{\theorembodyfont{\rmfamily}
}
{\theorembodyfont{\rmfamily}
}

\numberwithin{equation}{section}

\include{amsmath}
\begin{document}

\begin{title}
{\LARGE {\bf Jacobi equations and particle accelerator beam dynamics}}
\end{title}
\maketitle
\author{

\begin{center}

Ricardo Gallego Torrom\'e\\
Department of Physics, Lancaster University,\\
Lancaster, LA1 4YB \& The Cockcroft Institute, UK\footnote{{\bf email}: rgallegot@gmx.de.
{\bf Current address}: Instituto de Matem\'atica e Estat\'istica - USP, S\~ao Paulo, Brazil.}
\end{center}}

\begin{abstract}
A geometric formulation of the linear beam dynamics in accelerator physics is presented. In particular, it is proved that the linear transverse and longitudinal dynamics can be interpret geometrically as an approximation to the Jacobi equation of an affine {\it averaged Lorentz connection}. We introduce a specific notion reference trajectory as integral curves of the main velocity vector field. A perturbation caused by the statistical nature of the bunch of particles is considered.
\end{abstract}

\section{Introduction}

 The notion of {\it reference trajectory} is fundamental in particle accelerator beam dynamics \cite[Vol. I]{W}. The reference trajectory is a curve $X:{\bf I}\to {\bf R}^4$ associated to the motion of a bunch of particles in a particle accelerator. The reference trajectory is ideally the trajectory that one point charged particle will follow under the specifications of the initial conditions and external fields. The beam dynamics is adapted to the reference trajectory in such a way that the dynamics of particles composing each bunch are controlled by electric and magnetic fields, theoretically along trajectories near to the reference trajectory. Indeed,  beam parameters like {\it emittance} or {\it dispersion} and some observable quantities are referred to the motion relative to the reference trajectory.

However, there is no theoretical guarantee that one can associate to the reference trajectory a {\it real trajectory} of a charged point particle or the {\it real collective motion} of the bunch. This is because the observation of the motion of the bunch is indirect. Therefore, there is the assumption that the {reference trajectory} is related with a physical trajectory of the bunch of particles, although one does not have theoretical proof of that fact.

In this paper we introduce a theory where the reference trajectory is linked with an {\it observable quantity} that satisfies a geodesic differential equation. In particular, given a bunch of particles described by the kinetic function $f(x,y)$ \cite{Ande, D, Ehlers}, we consider the {\it averaged Lorentz dynamics} and the corresponding geodesic flow \cite{Ricardo09}. It determines trajectories associated with the {\it averaged velocity field} of the kinetic model, that we promote to be the expected velocity field.

The theory contains two different kind of elements. The first one is the $one$-particle distribution, that one takes as give  from kinetic model. Although we will assume that $f(x,y)$ is a solution of the Maxwell-Vlasov system or easier, the Vlasov's equation, the present paper only assumes that the distribution exists and is (weakly) smooth. We assume that such distribution function is given. The second kind of elements are from geometric nature, and related with the geometric theory of differential equations. In this paper we concentrate in the second aspect. We present two results. The first is that under some conditions, the standard {\it linear accelerator particle beam dynamics} corresponds to the Jacobi equation of the averaged Lorentz connection. Secondly, we present an observable which depends on the collective nature of the bunch of particles.

This work is organized as follows. In {\it section 2} we introduce the Lorentz connection and the averaged Lorentz connection. In {\it section 3} we discuss briefly the Jacobi equation of the averaged Lorentz connection. In {\it section 4} we show that the linear transversal dynamics in accelerator physics corresponds with a convenient approximation for the Jacobi equation of the averaged Lorentz connection. In {\it section 5} we find an observable which is related with the composed nature of the bunch of particles in a beam. {\it Section 6} is devoted to the {\it longitudinal dynamics} from the Jacobi equation point of view. Finally, a short discussion and perspectives for future work are briefly commented. 
\section{The Lorentz connection and the averaged Lorentz connection}
After introducing some notation, we define the Lorentz force connection and the averaged Lorentz connection and equations. We describe how the Lorentz force equation and the averaged Lorentz force equation are related.
\subsection{Linear connections on pull-back bundles}
Let $({\bf M},\eta)$ be the Minkowski space-time, {\bf TM} the tangent bundle, ${\bf N}=\,{\bf TM}\setminus\{0\}$ and ${\bf \Sigma}$ the unit tangent hyperboloid.
The pull-back bundle $\pi^* {\bf TM}\longrightarrow {\bf N}$ of the tangent bundle
{\bf TM} is the maximal sub-manifold of the cartesian product
${\bf N}\times{\bf TM} $ such that the following equivalence
relation holds: for every $u\in {\bf \Sigma} $ and $(u,\xi) \in \pi^{-1}
_1 (u)$, $ (u,\xi)\in {\bf \pi^* TM}$ {iff} $\pi \circ\pi
_2(u,\xi)=\pi(u)$, where the projection on the first and second factors
are denoted by
$\pi _1:\pi ^* {\bf TM}\longrightarrow {\bf N},\quad
(u,\xi)\longrightarrow u,\quad
\pi _2 :\pi ^* {\bf TM}\longrightarrow {\bf TM},\quad
(u,\xi)\longrightarrow \xi$.

A semi-spray vector field on ${\bf TN}$, written in local coordinates on {\bf N} $({\bf U},(x,y))$ by
  \begin{align}
  \frac{d y^i}{dt}=\,G^i(x,y),\quad y^i=\frac{dx}{dt}
  \end{align}
  has associated a Hessian of the form
\begin{align}
\Gamma^i \,_{jk}(x,y):=\frac{1}{2}\, \frac{\partial^2
G^i(x,y)}{\partial y^j \partial y^k}.
\label{equaciondecoeficientes de conexion}
\end{align}
The proof of the following result can be found in \cite{Ricardo09},
\begin{proposicion}
Given a semi-spray $G^i(x,y)$ there is defined a connection on $\pi^*{\bf TM}$ determined by the relations
\begin{align}
\nabla_{\frac{\delta}{\delta x^j}} \pi^* Z:=
\,^{\chi}\Gamma(x,y)^i\,_{jk}\,Z^k\,\pi^*e_i,\quad\quad \nabla_{V}
\pi^* Z:=0,\quad V\in \mathcal{V},\,Z\in \,\Gamma{\bf TM},
\end{align}
where $\{\pi^*e_i, \,i=0,1,2,3\,\}$ is a local frame for sections
$\Gamma(\pi^*{\bf TM})$.
\label{proposicionsobreconexiones}
\end{proposicion}
\subsection{The Lorentz connection}
The Lorentz force equation is the second order differential equation
\begin{align}
\frac{d^2 x^i}{dt^2} +\, ^{\eta}\Gamma^i\,_{jk}\,\frac{d
x^j}{dt}\frac{d x^k}{dt} +\eta^{ij}F_{jk} \frac{d
x^k}{dt}\sqrt{\eta(\frac{d x}{dt},\frac{d
x}{dt})}=0,\quad i,j,k=0,1,2,3,
\label{Lorentzequation}
\end{align}
for $t\in {\bf I}=[a,b]$, $^{\eta}\Gamma^i\,_{jk}(x)$ are the coefficients of the
Levi-Civita connection $^{\eta}\nabla$ of $\eta$ and $F$ is the Faraday tensor.

If we apply the formula (\ref{equaciondecoeficientes de conexion}) to the semi-spray associated with the Lorentz force, we obtain the connection coefficients \cite{Ricardo09}
\begin{align}
^L\Gamma^i\,_{jk} :=\, ^{\eta}\Gamma^i\,_{jk}+ ({\bf
F}^i\,_{j}\frac{1}{2}y^m\eta_{mk}+ {\bf
F}^i\,_{k}\frac{1}{2}y^m \eta_{mj})
+{\bf F}^i\, _m\,\frac{1}{2}\big(
{y^m}\eta_{jk}-\eta_{js}
\eta_{kl}\,y^m y^s y^l\,\big).
\label{gammalorentz}
\end{align}
 and one can define the corresponding connection by {\it proposition} \ref{proposicionsobreconexiones}.
It turns out that (see  \cite{Ricardo09}) the following proposition holds:
\begin{proposicion}
Let ${\bf M}$, ${\bf \Sigma}$, $\pi^*{\bf TM}$ and $^L\nabla$ be as
before. Then the auto-parallel curves of $^L\nabla$ are the solutions
 of the Lorentz force equation.
\end{proposicion}

\subsection{The averaged Lorentz connection}
Given the Lorentz connection $^L\nabla$ on the bundle $\pi^*{\bf
TM}\longrightarrow {\bf N}$ we associate an affine connection averaging on the unit tangent hyperboloid ${\bf \Sigma}$.
In order to calculate average of observables, we use the measure
\begin{displaymath}
dvol(x,y)=\sqrt{det\,{\eta}}\,\frac{1}{y^0}\,dy^1\wedge\cdot\cdot\cdot dy^{n-1},\quad y^0=y^0(x^0,...,x^{3},y^1,...,y^{3}).
\end{displaymath}
The function $f(x,y)$ is a (weak) solution of a one particle kinetic model. It can be associated with solutions of the Maxwell-Vlasov system or alternative dynamical models \cite{Ande, Ehlers}. 
Using such measure, one can prove that \cite{Ricardo09}:
\begin{proposicion}
The averaged connection of the Lorentz connection $^L\nabla$
is an affine, symmetric connection $\langle \,^L\nabla\rangle$ on  ${\bf M}$ with connection coefficients
\begin{align}
\langle\,^L\Gamma^i\,_{jk} \rangle:=\, ^{\eta}\Gamma^i\,_{jk}+ ({\bf
F}^i\,_{j}\langle\frac{1}{2}y^m\eta_{jk}\rangle\eta_{mk}+ {\bf
F}^i\,_{k}\langle\frac{1}{2}y^m\rangle\eta_{mj})
+{\bf F}^i\, _m\,\frac{1}{2}\big(
\langle{y^m}\rangle\,\eta_{jk}-\eta_{js}
\eta_{kl}\langle\,y^m y^s y^l\rangle\,\big).
\label{gammalorentzpromediado}
\end{align}
The moments are defined by the integrals
\begin{align*}
& {vol({\bf \Sigma}_x)}  :=\int _{{\bf \Sigma}_x}
f(x,y)\,dvol(x,y),\quad \langle y^i\rangle :=\frac{1}{vol({\bf \Sigma}_x)}\int _{{\bf \Sigma}_x} y^i
f(x,y)\,dvol(x,y),\\
&\langle y^m y^s y^l\rangle :=\frac{1}{vol({\bf \Sigma}_x)}\int _{{\bf \Sigma}_x} y^m y^s y^l
f(x,y)\,dvol(x,y).
\end{align*}
\end{proposicion}
\subsection{Comparison between the geodesics of $\,^L\nabla$ and $\langle\,^L\nabla\rangle$}
We define the energy function $E$ of the one particle distribution function $f(x,y)$ to be the real function
\begin{equation}
E:{\bf M}\longrightarrow {\bf R},\quad
x\mapsto E(x):=inf\{ y^0,\, y\in supp(f_x)\},
\end{equation}
where $y^0$ is the $0$-component of the tangent velocity vector,
 measured in the laboratory coordinate frame. The function $\alpha(x)$ is the diameter of the support of $f_x$; the $\alpha$ is defined as $\alpha=\,supp\{\alpha(x),\,x\in\,{\bf M}\}$. The parameter $s$ is the coordinate time measured in the laboratory frame.

The relation between the geodesics of $^L\nabla$ and the geodesics of $\langle\,^L\nabla\rangle$ by the following result \cite{Ricardo09, GT2}
\begin{teorema}
Let $({\bf M},\eta, [A])$ be a semi-Randers space. Let us assume that
\begin{enumerate}

\item The auto-parallel curves of unit velocity of the connections $^L\nabla$ and
 $\langle\,^L\nabla\rangle $ are defined on $t\in\,{\bf I}$.

\item The ultra-relativistic limit holds: the energy function of the beam $E(x(s))$ is much larger than the rest mass of the particles, $E(x(s))\rangle  \rangle  1$ for all $t\in {\bf I }$.

\item The distribution function is narrow $f(x,y)=0$ in the sense that ${\alpha}<<1$ for all $x\in {\bf M}$.

\item The support of the distribution function $f(x,y)$ is invariant under the flow of the Lorentz force equation.

\item The change in the energy function is adiabatic in the sense that $\frac{d}{ds} log E(x(s))<<1$.
\end{enumerate}
Then for the same arbitrary initial condition $(x(0),\dot{x}(0))$, the solutions of the equations
\begin{displaymath}
^L\nabla_{\dot{x}} \dot{x}=0,\, \quad \langle\,^L\nabla\rangle_{\dot{\tilde{x}}} \dot{\tilde{x}}=0
\end{displaymath}
differ in such a way that
\begin{equation}
\|\tilde{x}(s)-\, x(s)\|\leq\, 2\big(C(x)\|{\bf F}\|(x)\,+E^{-1}(s)\,C^2_2(x)(1+B_2(x){\alpha})\big)
{\alpha}^2\,E^{-2}(s)\,s^2,
\end{equation}
where the functions $C(x)$, $C_2(x)$ and $B_2(x)$ are bounded by constants of order $1$ and the distances are measured in the laboratory frame.
\label{teoremadecomparaciondegeodesicas}
\end{teorema}
 One of the consequences of this result is that in the ultra-relativistic limit, and for narrow distributions, the solutions of the Lorentz force equation can be approximate by geodesics of the averaged Lorentz connection $\langle\,^L\nabla\rangle$,
 \begin{align}
 \langle \,^L\nabla_{\dot{z}}\rangle\,\dot{z}=0.
 \label{averagedgeodesicequation}
 \end{align}
 The integral curves of the {\it mean velocity field}  $\tilde{V}(x)\simeq \langle y\rangle(x)$ are approximate solutions of equation \ref{averagedgeodesicequation} 
 (see for instance \cite{GT1}),
 \begin{align}
 \|\langle\, ^L\nabla \rangle_{\tilde{V}} \tilde{V}(x)\|\leq C({\bf K})\cdot {\alpha}^2\,+ O(\alpha^3).
 \label{equationfor<v>}
 \end{align}
 where C({\bf K }) is a constant. Therefore, the geodesics of $^L\nabla$ can be approximated by the integral curves of the mean vector field $\langle y\rangle(x)$ as equation (\ref{equationfor<v>}) shows\footnote{In equilibrium, one makes the assumption that $\langle y\rangle$ is equal to the expected velocity field.}.

\section{The Jacobi equation of the averaged Lorentz dynamics}
We consider the Jacobi equation of a general affine, torsion free connection $\nabla$. 
The Jacobi equation is a linear differential equations describing the relative behavior of nearby geodesics.
If {\it the central geodesic} is $X(t)$ and $x(t)=\xi(t)+X(t)$  neighborhood geodesic is $x(t)=\xi(t)+X(t)$, then the Jacobi equation for $\nabla$ is the linear, ordinary differential equation
\begin{align}
\nabla_{\dot{X}}\nabla_{\dot{X}}\,(\xi)\,+ R(\xi,\dot{X})\cdot \dot{X}=0,
\label{Jacobiequation}
\end{align}
where $R(\xi,\dot{X})$ is the curvature endomorphism of the plane generated by $\{\xi,\dot{X}\}$ \cite{Hicks}.
The Jacobi equation can be re-written in local coordinates as
\begin{align}
\frac{d^2 \xi^{i}}{dt^2}+2\Gamma^{i}\,_{j
k}(X(t))\frac{d\xi^j}{dt}\frac{dX^k}{d t}+\xi^{l}\partial_{l}
\Gamma^{j}\,_{jk}(X)\frac{d
X^j}{d t}\frac{dX^k}{d t}=0.
\label{geodesicdeviationequation}
\end{align}
The averaged Lorentz connection $\langle\,^L\nabla\rangle$ is an affine connection on
the tangent bundle {\bf TM}. Therefore, we can apply the standard
Jacobi equation to the averaged Lorentz connection,
\begin{align*}
0 &= \frac{d^2 \xi^i}{dt^2}+
2\frac{d\xi^j}{dt}\frac{dX^k}{dt}\Big(\frac{1}{2}({\bf
F}^i\,_{j}\langle y^m\rangle\eta_{mk}+ {\bf F}^i\,_{k}\langle y^m\rangle \eta_{mj})\\
& +{\bf F}^i\, _m\big( {\langle y^m\rangle }\eta_{jk}-\eta_{js} \eta_{kl}\langle y^m y^s
y^l\rangle \big)\Big) \\
& +2\xi^l \partial_l \Big(\frac{1}{2}({\bf
F}^i\,_{j}\langle y^m\rangle \eta_{mk}+ {\bf F}^i\,_{k}\langle y^m \rangle \eta_{mj}) \\
& +{\bf F}^i\, _m\big( {\langle y^m\rangle }\eta_{jk}-\eta_{js} \eta_{kl}\langle y^m y^s
y^l\rangle \big)\Big)\frac{dX^j}{dt}\frac{dX^k}{dt}
\end{align*}
\begin{align}
+\big(\,^{\eta}\Gamma^i\,_{jk}+\xi^l\partial_l
\,^{\eta}\Gamma^i_{jk}\big)
\big(\frac{dX^j}{dt}\frac{dX^k}{dt}+2\frac{dX^j}{ds}
\frac{d\xi^k}{dt}\big).
\label{jacobiequationaveragedconnection}
\end{align}
From the form of the system of differential equations (\ref{jacobiequationaveragedconnection})  it is clear the following,
\begin{enumerate}
\item There is a term representing the {\it inertial force}:
\begin{equation}
\mathcal{A}_I:=\big(\,^{\eta}\Gamma^i\,_{jk}+\xi^l\partial_l
\,^{\eta}\Gamma^i\,_{jk}\big)
\big(\frac{dX^j}{dt}\frac{dX^k}{dt}+2\frac{dX^j}{dt}\frac{d\xi^k}{dt}\big).
\end{equation}
$\mathcal{A_I}$ is universal, in the sense that it is independent of the particle mass.

\item $\mathcal{A_I}$ can depend on the electromagnetic field,
since $\frac{dX^j}{dt}$ can depend implicitly on the electromagnetic field
when defining the reference trajectory. The typical example is the reference orbit in a {\it betatron accelerator} \cite[Vol. I, chapter 3]{W}.
\end{enumerate}
\section{Transversal beam dynamics from the
Jacobi equation of the averaged Lorentz connection}
Let us consider the {\it laboratory coordinate frame}. Then a convenient global coordinate
system $(t,x^1,x^2,x^3)$ is defined as follows. The parameter $t$ is the proper time of the reference
trajectory, the coordinate $x^2$ is the Euclidean length measured in the laboratory frame, and $x^2$, $x^3$ are the orthogonal coordinates of a point in directions orthogonal to $\frac{\partial}{\partial x^2}$. Given the reference geodesic $X: {\bf I} \to {\bf M}$ is very useful to consider the relative coordinates
 $(x^1-X^1, x^3-X^3)$ respect the reference trajectory (transverse degrees of freedom). In this way, $(\xi^1,x^2,\xi^3)$ coincide with the variables used in beam dynamics \cite[Chapter 4]{W}.
\subsection{Transverse dynamics and the Jacobi equation of the averaged Lorentz connection}

 For the analysis of the {\it transverse dynamics} we  assume that
the difference $\frac{dx^2}{dt}-\frac{dX^2}{dt}$ is constant and that the external
electromagnetic fields are static magnetic fields in Minkowski space.
We will only consider the lower order terms in the degree $a+b+c$ of the monomials $\xi^a(\frac{d\xi}{dt})^b\epsilon^c$ appearing in the corresponding expressions, with $\xi=x(t)-X(t),\quad \epsilon :=\langle y\rangle  -\frac{dX}{dt}$.

First we linearize the equations with respect to the degree defined
by the vector fields along the central geodesic $\xi$, its derivatives and $\epsilon$,
\begin{align*}
0 & =\frac{d^2 \xi^i}{dt^2}+
2\frac{d\xi^j}{dt}\frac{dX^k}{dt}\Big(\frac{1}{2}({\bf
F}^i\,_{j}\langle y^m\rangle  \eta_{mk}+ {\bf F}^i\,_{k}\langle y^m\rangle  \eta_{mj})\Big) \\
& +2\xi^l
\partial_l \Big(\frac{1}{2}({\bf F}^i\,_{j}\langle y^m\rangle  \eta_{mk}+ {\bf
F}^i\,_{k}\langle y^m\rangle  \eta_{mj})\Big)\cdot\frac{dX^j}{dt}\frac{dX^k}{dt}\\
& +\big(\,^{\eta}\Gamma^i\,_{jk}+\xi^l\partial_l
\,^{\eta}\Gamma^i\,_{jk}\big)
\big(\frac{dX^j}{dt}\frac{dX^k}{dt}+2\frac{dX^j}{dt}
\frac{d\xi^k}{dt}\big), \quad i,j,k,m=0,1,2,3.
\end{align*}
Since $\epsilon$ is small, we can replace
$\langle y\rangle  \longrightarrow \frac{dX}{ds}$, obtaining the expression
\begin{align*}
0 & =\frac{d^2 \xi^i}{dt^2}+
2\frac{d\xi^j}{dt}\frac{dX^k}{dt}\Big(\frac{1}{2}({\bf F}^i\,_{j}
\frac{dX^m}{dt}\eta_{mk}+{\bf F}^i\,_{k}\frac{dX^m}{dt}\eta_{mj})\Big)\\
& +2\xi^l
\partial_l \Big(\frac{1}{2}({\bf F}^i\,_{j}\frac{dX^m}{dt}\eta_{mk}+ {\bf
F}^i\,_{k}\frac{dX^m}{dt}\eta_{mj})\Big)\cdot\frac{dX^j}{dt}\frac{dX^k}{dt}\\
& +\big(\,^{\eta}\Gamma^i\,_{jk}+\xi^l\partial_l
\,^{\eta}\Gamma^i\,_{jk}\big)
\big(\frac{dX^j}{dt}\frac{dX^k}{dt}+2\frac{dX^j}{dt}
\frac{d\xi^k}{dt}\big).
\end{align*}
The condition of transversal dynamics is
$\frac{d \xi^j}{dt} \frac{dX_j}{dt}$ is second order in $\xi$ and its derivatives. Suppressing this term we obtain,
\begin{align*}
0 & =\frac{d^2 \xi^i}{dt^2}+
2\frac{d\xi^j}{dt}\frac{dX^k}{dt}\Big(\frac{1}{2}{\bf F}^i\,_{j}
\frac{dX^m}{dt}\eta_{mk}\Big)\,+ 2\xi^l
\partial_l \Big(\frac{1}{2}{\bf
F}^i\,_{j}\frac{dX^m}{dt}\eta_{mk}\Big)\frac{dX^j}{dt}\frac{dX^k}{dt}\\
& +\big(\,^{\eta}\Gamma^i\,_{jk}+\xi^l\partial_l
\,^{\eta}\Gamma^i\,_{jk}\big)
\big(\frac{dX^j}{dt}\frac{dX^k}{dt}+2\frac{dX^j}{dt}\frac{d\xi^k}{dt}\big).
\end{align*}
Therefore, at first order the differential equations are
\begin{align*}
0 &=\frac{d^2 \xi^i}{dt^2}+ \frac{d\xi^j}{dt}{\bf F}^i\,_{j} + \frac{d
X^j}{dt}\xi^l
\partial_l {\bf F}^i\,_{j}
+\,^{\eta}\Gamma^i\,_{jk}\frac{dX^j}{dt}\frac{dX^k}{dt} +\xi^l\partial_l
\,^{\eta}\Gamma^i\,_{jk}\frac{dX^j}{dt}\frac{dX^k}{dt}+
2\,^{\eta}\Gamma^i\,_{jk}\frac{dX^j}{dt}\frac{d\xi^k}{dt}.
\end{align*}
In the transverse dynamics, if there is not {\it dispersion} one has that
$\frac{d\xi^j}{dt}{\bf F}^i\,_j=0$, if there is not {\it dispersion}. Hence, the differential equations in this regime are
\begin{equation}
0=\frac{d^2 \xi^i}{dt^2}+ \frac{d\xi^j}{dt}{\bf F}^i\,_{j}
+\,^{\eta}\Gamma^i\,_{jk}\frac{dX^j}{dt}\frac{dX^k}{dt}+\xi^l\partial_l
\,^{\eta}\Gamma^i\,_{jk}\frac{dX^j}{dt}\frac{dX^k}{dt}+
2\,^{\eta}\Gamma^i\,_{jk}\frac{dX^j}{dt}\frac{d\xi^k}{dt}.
\label{transversecovariantdynamics}
\end{equation}
The last term in (\ref{transversecovariantdynamics}) is an {\it inertial term}. We assume that $\xi$ is small compared with
 the curvature radius $\rho$ of the central geodesic. Then the inertial terms are
\begin{align*}
& 0=\big(\,^{\eta}\Gamma^0\,_{jk}+\xi^l\partial_l
\,^{\eta}\Gamma^0\,_{jk}\big)
\big(\frac{dX^j}{dt}\frac{dX^k}{dt}+2\frac{dX^j}{dt}\frac{d\xi^k}{dt}\big),\\
& 0=\big(\,^{\eta}\Gamma^3\,_{jk}+\xi^l\partial_l
\,^{\eta}\Gamma^3\,_{jk}\big)
\big(\frac{dX^j}{dt}\frac{dX^k}{dt}d+2\frac{dX^j}{dt}
\frac{d\xi^k}{dt}\big),\\
&0=\big(\,^{\eta}\Gamma^2\,_{jk}+\xi^l\partial_l
\,^{\eta}\Gamma^2\,_{jk}\big)
\big(\frac{dX^j}{dt}\frac{dX^k}{dt}+2\frac{dX^j}{dt}\frac{d\xi^k}{dt}\big).\\
&
\big(\frac{d\vec{X}}{dt}\big)^2\,\frac{\xi^1}{\rho^2}=\big(\,^{\eta}\Gamma^1\,_{jk}+\xi^l\partial_l
\,^{\eta}\Gamma^1_{jk}\big)
\big(\frac{dX^j}{dt}\frac{dX^k}{dt}+2\frac{dX^j}{dt}\frac{d\xi^k}{dt}\big),
\end{align*}
The last term in the fourth equation corresponds to the {\it relative centripetal force}.
\subsection{Examples of transverse linear dynamics}
We study some examples of transverse linear dynamics using the {\it linearized version} of the averaged Jacobi equation.
\begin{enumerate}
\item {\bf Motion in a normal magnetic dipole}

The reference frame is the laboratory frame.
In this case the electromagnetic field is given by the expression
\begin{displaymath}
{\bf F}=\left(%
\begin{array}{cccc}
  0 & 0 & 0 & 0 \\
  0 & 0 & b_0 & 0 \\
  0 & -b_0 & 0 & 0 \\
  0 & 0 & 0 & 0 \\
\end{array}%
\right)
\end{displaymath}
where $b_0$ is the dipole strength. Since the magnetic field is constant, $\xi^l\partial_l {\bf F}^i\,_j =0$. Therefore, the equations
of motion for the transverse degrees of freedom $(\xi^1,\xi^3)$ are
\begin{displaymath}
\frac{d^2 \xi^1}{dt^2}
+\big(\frac{d\vec{X}}{dt}\big)^2\,(\frac{\xi^1}{\rho^2})=0,\quad
\frac{d^2 \xi^3}{dt^2}=0.
\end{displaymath}
One can change the parameter of the curves by $t\to x^2=l$. Then one has that
\begin{displaymath}
\frac{d^2 \xi^1}{dl^2}+\frac{\xi^1}{\rho^2}=0,\quad \frac{d^2
\xi^3}{dl^2}=0.
\end{displaymath}
These are the standard equations for a normal dipole.

\item {\bf Motion in a normal quadrupole field combined with a dipole}

In this case the electromagnetic field has the form
\begin{displaymath}
{\bf F}(x)=\left(%
\begin{array}{cccc}
  0 & 0 & 0 & 0 \\
  0 & 0 & b_0-b_1 \xi^1 & 0\\
  0 & -b_0+b_1 \xi^1 & 0 & b_1 \xi^3 \\
  0 & 0 & -b_1 \xi^3 & 0\\
\end{array}%
\right)
\end{displaymath}
The Jacobi equation reduces to
\begin{displaymath}
\frac{d^2 \xi^1}{dt^2}-
\frac{d X^j}{dt}\xi^l
\partial_{l} {\bf F}^1\,_{j}
+\big(\frac{d\vec{X}}{dt}\big)^2\,(\frac{\xi^1}{\rho^2})=0,\quad
\frac{d^2 \xi^3}{dt^2} +
\frac{dX^j}{dt}\xi^l
\partial_l {\bf F}^3\,_{j}=0.
\end{displaymath}
Let us consider the contributions from $ \frac{d\xi^j}{dt}
\xi^l
\partial_l {\bf F}^1\,_{j}$ and $ \frac{d\xi^j}{dt} \xi^l
\partial_l {\bf F}^3\,_{j}$. Using Euler's theorem on homogenous
functions one gets the relations:
\begin{displaymath}
\frac{d X^j}{dt}\xi^l
\partial_l {\bf F}^3\,_{j}=\frac{d X^j}{dt}{\bf F}^3_{j}|_{\xi=0}.
\end{displaymath}
Then the differential equations are
\begin{displaymath}
\frac{d^2 \xi^1}{dt^2}-\xi^1 b_1
+\big(\frac{d\vec{X}}{dt}\big)^2\,(\frac{\xi^1}{\rho}^2)=0,\quad
\frac{d^2 \xi^3}{dt^2}+\xi^3 b_1=0.
\end{displaymath}
These are the equations of the linear transverse dynamics in
quadrupoles combined with magnetic dipole field. The solutions are parameterized by the proper time $t$. If we use the Euclidean length $l$, the equations
are
\begin{equation}
\frac{d^2\xi^1}{dl^2}-\xi^1\,b_1 +\frac{\xi^1}{\rho^2}=0, \quad
\frac{d^2 \xi^3}{dl^2}+\xi^3\,b_0=0.
\end{equation}
These equations coincide with the standard equations in beam dynamics \cite[Vol.I, chapter 4]{W}.

\item {\bf Motion in a normal dipole combined with a 45 degrees quadrupole}

In this case the electromagnetic field is
\begin{displaymath}
{\bf F}=\left(%
\begin{array}{cccc}
  0 & 0 & 0 & 0 \\
  0 & 0 & b_0+b_1 \xi^3 & 0\\
  0 & -b_0-b_1 \xi^3 & 0 & b_1 \xi^1 \\
  0 & 0 & -b_1 \xi^1 & 0\\
\end{array}%
\right)
\end{displaymath}
Following the same procedure as before, we get the Jacobi equations
\begin{equation}
\frac{d^2\xi^1}{dl^2}+\xi^1\,b_1 +\frac{\xi^1}{\rho^2}=0, \quad
\frac{d^2 \xi^3}{dl^2}-\xi^3\,b_0=0.
\end{equation}
\end{enumerate}
The above examples show how the linear transverse dynamics can be
obtained from the Jacobi equation of the averaged Lorentz connection.

\section{A perturbative collective dynamical effect}

We will consider the perturbation in the equation of a point charged particle caused by the
fact that the bunch of particles is composed by a collection of many identical particles.

\subsection{Calculation of the dispersion function in beam dynamics}

We follow the formalism developed in \cite[Chapter 4]{W} for the treatment
of linear perturbations and dispersion. However, we
maintain the proper time $t$ as the parameter of the curves.

As we showed in the preceding {\it section}, the transverse dynamics is determined by equations of the form
\begin{align}
\frac{d^2 u}{dt^2} \, +K(t)u=0
\label{hillequation}
\end{align}
The general solution of (\ref{hillequation}) is
\begin{displaymath}
u(t)=C(t)u_0\,+S(t)\frac{d u_0}{dt},\quad \frac{d u}{dt}(t)=\frac{d C}{dt}(t)u_0+\frac{d S}{dt}(t)\frac{d u_0}{dt},
\end{displaymath}
with initial conditions
\begin{align*}
C(0)=1,\quad \frac{d C}{dt}(0)=0;\quad S(0)=0,\quad \frac{d S}{dt}(0)=1
\end{align*}
 for arbitrary initial values $u_0$ and
$\frac{d u_0}{dt}$. The functions $C(t)$ and $S(t)$ satisfy
\begin{displaymath}
\frac{d^2 C}{dt^2}(t)+K(t)S(t)=0,\quad \frac{d^2 S}{dt^2}(t)+K(t)S(t)=0.
\end{displaymath}
The perturbed equation has the form:
\begin{equation}
\frac{d^2 u}{dt^2}(t) \, +K(t)u(t)p(t).
\label{perturbedequationforu}
\end{equation}
A particular solution for (\ref{perturbedequationforu}) is
\begin{equation}
P(t)=\int^{t}_0 p(\tilde{t})G(t,\tilde{t})d\tilde{t},
\label{particularsolutiooftheperturbedequation}
\end{equation}
where $G(t,\tilde{t})$ is the Green function associated to the differential equation (\ref{perturbedequationforu}).
One can prove that in the absence of dissipative forces (that is, which do not depend on the velocity of the particle), the Green
function of the differential equation is
\begin{equation}
G(t,\tilde{t})=S(t)C(\tilde{t})-C(t)S(\tilde{t}).
\end{equation}
Therefore, the general solution for the equation (\ref{perturbedequationforu}) is
\begin{equation}
u(t)=a_u\,C(t)+b_u\,S(t)\,+P(t).
\end{equation}
 This solution breaks down if there are strong loses of energy by synchrotron radiation or other dissipative effects.

 If the components of a  bunch do not have the same energy, one obtains for the transverse degrees of
freedom the following differential equation \cite[pg 109]{W},
\begin{align*}
\frac{d^2 u}{dt^2} \, +K(t)u\,=\frac{1}{\rho_0}(t)\delta_u, \quad
\Delta=\frac{\delta p}{p_0},\quad \delta p=\sqrt{(\delta
p_1)^2\,+(\delta p_2)^2\, +(\delta p_3)^2}.
\end{align*}
We assign  to $\delta p$  the maximal
value of $\{\|\vec{\xi(X(t))}\|\}$, using the Euclidean metric defined on the laboratory coordinate system. Then
the general solution for $u$ is linear in the perturbation,
\begin{displaymath}
u(t)=a_u\,x(t)+b_u\,S(t)\,+\Delta\,D(t):=a_u\,x(t)+b_u\,S(t)\,
+Off_u(t).
\end{displaymath}
where $a_u$ and $b_u$ are constants that depend on the initial values and $D(t)$ is the dispersion function as in \cite[Chapter 4]{W}.

\subsection{An observable associated with the composed nature of the bunch of particles}
We have shown in the previous section that at first order and when we take the approximation $\langle y\rangle  \to
\frac{d X}{dt}$, the differential
equation for the transverse motion is the Jacobi equation of the
averaged connection. Therefore we can consider the terms on $\epsilon^k$
in the averaged Jacobi equation as a perturbation and apply the method of the Green function.

The off-set function is defined as
\begin{displaymath}
Off^{1,3}_{\xi}(t)=u^{1,3}(t)-a_{1,3}\,C^{1,3}(t)-\,
b_{1,3}\,S^{1,3}(t),
\end{displaymath}
the super-index refers to the transverse directions ($\frac{\partial}{\partial x^1}$ or $\frac{\partial}{\partial x^3}$ in the laboratory frame defined previously.
Using the corresponding Green function we obtain
\begin{align}
Off^{1,3}_{\xi}(t) = \int^{t} _0\, p^{1,3}(t)G_u(t,\tilde{t})d\tilde{t}.
\label{definitionoffset}
\end{align}
The perturbation $p^{1,3}(t)$ is in this case defined by all the terms of the averaged
Jacobi equation which are not contained in the linearized equation respect to the degree $(a+b+c)$. Therefore let us re-write the
Jacobi equation of the averaged connection. Using $\epsilon_k\,=\,\langle y_k\rangle\,-\frac{dX_k}{d\tilde{t}}$ we get
\begin{align*}
Off^{1,3}_{\xi}(t) & = \int^{t}_0\,d\tilde{t}\, 2\,\frac{d
\xi^j}{d\tilde{t}}(\tilde{t})\cdot
\frac{dX^k}{d\tilde{\tilde{t}}}\Big(\frac{1}{2}\big(\,{\bf
F}^{1,3}\,_j(\tilde{t})\,\epsilon_k(\tilde{t})\,+{\bf F}^{1,3}\,_k(\tilde{t})\, \epsilon_j(\tilde{t})\,\big)\\
& + \frac{dX^j}{d\tilde{t}}\frac{dX^k}{d\tilde{t}}\Big( \,{\bf F}^{1,3}\,_m(\tilde{t})\, \big(\,
\langle y^m\rangle (\tilde{t})\eta_{jk}\,-\langle y^m y^a y^k\rangle(\tilde{t})\eta_{ja}\eta_{lk}\big)+
\end{align*}
\begin{equation}
+\xi^l\partial_l \,\big(\,{\bf F}^{1,3}\,_m(\tilde{t})\, \big(\,
\langle y^m\rangle (\tilde{t})\eta_{jk}\,-\langle y^m y^a y^k\rangle (\tilde{t})\eta_{ja}\eta_{lk}\big)\Big)\Big).
\label{integrodifferentialequation}
\end{equation}
In the integrand of equation (\ref{integrodifferentialequation}) we can make the substitution
\begin{displaymath}
\frac{d \xi^{1,3}}{d\tilde{t}}(\tilde{t})\longrightarrow
\big(a_{1,3}\,\frac{d C}{dt}^{1,3}(\tilde{t})+b_{1,3}\,\frac{d S}{dt}^{1,3}(\tilde{t})+
\frac{Off^{1,3}_{\xi}(\tilde{t})}{dt}\big).
\end{displaymath}
We can also consider the derivatives in the longitudinal and
temporal directions.  With a convenient choice
of the coefficients $a_{0,2}$ and $b_{0,2}$. Then we can write
\begin{displaymath}
\frac{d \xi^{j}}{d\tilde{t}}(\tilde{t})\longrightarrow
\big(a_j\,\frac{d C}{dt}^j(\tilde{t})+b_j\,\frac{d S}{dt}^j(\tilde{t})+\,\frac{Off^j_{\xi}(\tilde{t})}{dt}\big),\quad
j=0,1,2,3.
\end{displaymath}
In this expression repeated indices are not summed.
For the transverse degrees of freedom, the unperturbed solutions
are the same as before.

For the longitudinal $j=2$ and temporal $j=0$ degrees of freedom,
one gets the following relations by comparison with the Jacobi
equation,
\begin{align*}
&\frac{d \xi^{2}}{d\tilde{t}}(\tilde{t})\longrightarrow
\big(a_2\,\frac{d C}{dt}^2(\tilde{t})+b_2\,\frac{d S}{dt}^2(\tilde{t})+\frac{Off^2_u(\tilde{t})}{dt}\big),
\\
&\frac{d \xi^{0}}{d\tilde{t}}(\tilde{t})\longrightarrow
\big(a_0\,\frac{d C}{dt}^0(\tilde{t})+b_0\,\frac{d S}{dt}^0(\tilde{t})+\frac{Off^0_u(\tilde{t})}{dt}\big).
\end{align*}
Let us consider the regime where $Off^0_{\xi}=Off^2_{\xi} =0,\,\, \forall u$.
Then
\begin{align*}
Off^{1,3}_{\xi} (t) & = \int^{t}_0\,d\tilde{t}\,\Big(\sum^{3}_{j=0}
2\big(a_j\,\frac{d C}{dt}^{j}(\tilde{t})+b_j\,\frac{d S}{dt}^{j}(\tilde{t})+
Off^{j}_u(\tilde{t})'\big)(\tilde{t})\cdot
\frac{dX^k}{d\tilde{t}}\Big(\frac{1}{2}\big(\,{\bf
F}^{1,3}\,_j(\tilde{t})\,\epsilon_k(\tilde{t})\\
& + {\bf F}^{1,3}\,_k(\tilde{t})\, \epsilon_j(\tilde{t})\,\big)\,+
\frac{dX^j}{d\tilde{t}}\frac{dX^k}{d\tilde{t}}\Big({\bf F}^{1,3}\,_m(\tilde{t})\, \big(\,
\langle y^m\rangle  (\tilde{t})\eta_{jk}\,-\langle y^m y^a y^k\rangle  (\tilde{t})\eta_{ja}\eta_{lk}\big)\\
& +\big(a_l\,C^{l}(\tilde{t})+b_l\,S^{l}(\tilde{t})+
Off^{l}_u(\tilde{t})\big)\partial_l \,\big(\,{\bf F}^{1,3}\,_m\,
\big(\, \langle y^m\rangle  (\tilde{t})\eta_{jk}\,-\langle y^m y^a
y^l\rangle  (\tilde{t})\eta_{ja}\eta_{lk}\big)\Big)\Big).
\end{align*}
This is an integro-differential equation for $Off^{1,3}_{\xi}$. In {\it the Born approximation} one puts $Off^{l}_{\xi}(\tilde{t})=0$ in the integrand,
\begin{align*}
Off^{1,3}_{\xi}(t)& = \int^{t}_0\,d\tilde{t}\,\Big(
2\frac{d\xi^j}{d\tilde{t}}\cdot
\frac{dX^k}{d\tilde{t}}\Big(\frac{1}{2}\big(\,{\bf
F}^{1,3}\,_j(\tilde{t})\,\epsilon_k(\tilde{t})\, + {\bf F}^{1,3}\,_k(\tilde{t})\, \epsilon_j(\tilde{t})\,\big)\\
& +\frac{dX^j}{d\tilde{t}}\frac{dX^k}{d\tilde{t}}\Big( \,{\bf F}^{1,3}\,_m\, \big(\,
\langle y^m\rangle  (\tilde{t})\eta_{jk}\,-\langle y^m y^a y^k\rangle  (\tilde{t})\eta_{ja}\eta_{lk}\big)
\end{align*}
\begin{align}
+\xi^l\partial_l \,\big(\,{\bf F}^{1,3}\,_m(\tilde{t})\, \big(\,
\langle y^m\rangle  (\tilde{t})\eta_{jk}\,-\langle y^m y^a y^l\rangle  (\tilde{t})\eta_{ja}\eta_{lk}\big)\Big)\Big).
\end{align}
If the perturbation is constant along the reference trajectory, the term
containing derivatives are neglected and one obtains
\begin{displaymath}
Off^{1,3}_{\xi}(t) = \int^{t}_0\,d\tilde{t}\,\langle \Big(
2\frac{d\xi^j}{d\tilde{t}}\frac{dX^k}{d\tilde{t}}\cdot \big(\,{\bf
F}^{1,3}\,_j(\tilde{t})\,\epsilon_k (\tilde{t})+{\bf F}^{1,3}\,_k(\tilde{t})\, \epsilon_j(\tilde{t})\big)\rangle  \, +
\end{displaymath}
\begin{equation}
+ \frac{dX^j}{d\tilde{t}}\frac{dX^k}{d\tilde{t}}\Big( \,{\bf F}^{1,3}\,_m(\tilde{t})\, \big(\,
\langle y^m\rangle  (\tilde{t})\eta_{jk}\,-\langle y^m y^a y^k\rangle  (\tilde{t})\eta_{ja}\eta_{lk}\big)\Big)\Big).
\end{equation}
This expression depends on the particular solution $\frac{d\xi}{dt}(t)$. One way to eliminate such dependence is to consider the {\it average} of the above expression with $y=\frac{dx}{dt}$ for all possible geodesics (note that then $\xi_k=\epsilon_k$),
\begin{align*}
\langle Off^{1,3}_{\xi}\rangle  (t)  &= \int^{t}_0\,d\tilde{t}\,\Big(
\langle 2\frac{d\xi^j}{dt}(\tilde{t})\cdot
\frac{dX^k}{d\tilde{t}}\Big(\frac{1}{2}\big(\,{\bf
F}^{1,3}\,_j(\tilde{t})\,\xi_k(\tilde{t}) +{\bf F}^{1,3}\,_k(\tilde{t})\, \xi_j(\tilde{t})\big)\rangle\\
& + \frac{dX^j}{d\tilde{t}}\frac{dX^k}{d\tilde{t}}\Big( \,{\bf F}^{1,3}\,_m(\tilde{t})\, \big(\,
\langle y^m\rangle  (\tilde{t})\eta_{jk}\,-\langle y^m y^a y^k\rangle  (\tilde{t})\eta_{ja}\eta_{lk}\big)\Big)
\end{align*}
However, the average of the first integral is zero. Therefore.
\begin{align}
\langle Off^{1,3}_{\xi}\rangle  (t) \,=\int^{t}_0\,d\tilde{t}\,\Big(\frac{dX^j}{d\tilde{t}}\frac{dX^k}{d\tilde{t}}\Big( \,{\bf F}^{1,3}\,_m(\tilde{t})\, \big(\,
\langle y^m\rangle  (\tilde{t})\eta_{jk}\,-\langle y^m y^a y^k\rangle  (\tilde{t})\eta_{ja}\eta_{lk}\big)\Big).
\end{align}
In the case of a delta function distribution one has that $\langle Off^{1,3}_{\xi}\rangle  (t)=0$. Therefore, {\it the averaged off-set function} $\langle Off^{1,3}_{\xi}\rangle  (t)$ is a measurement of which good is the approximation of the standard beam dynamics as description of the dynamics of a system of particles.

The averaged off-set $\langle Off^{1,3}_{\xi}\rangle  (t) $  is determined by:
\begin{enumerate}
\item The reference trajectory $X(t)$, which is a geodesic of the averaged connection. This is theoretically fixed and corresponds to an observable quantity.

\item The external electromagnetic field ${\bf F}^{1,3}\,_m(X(t))$ along the reference trajectory.

\item The first and third momentum moments of the distribution function
$f(x,y)$ along the reference trajectory.
\end{enumerate}
This properties make the function $\langle Off^{1,3}_{\xi}\rangle  (t)$ to be an observable associated with the beam of particles. 

There is a dependence on how the averages are calculated. This depends on the specific kinetic model that we are considering. One this is fixed, the function $\langle Off^{1,3}_{\xi}\rangle  (t)$ is useful for two things, depending on the particular situation: if the kinetic model is well known, it can be used to test the dynamics of the accelerator; if the dynamics of the beam is well known, it can used to test the kinetic model used. It is also interesting that indeed it depends on only three moments of the distribution function.

\section{Longitudinal beam dynamics and corrections
from the Jacobi equation of the averaged Lorentz connection}
{\it Longitudinal dynamics} is the theory in beam dynamics that deal with the acceleration mechanisms \cite[Vol. I, chapter 8]{W}. We show that in both, the Jacobi equation of the averaged Lorentz dynamics provides equations for longitudinal dynamics, even if in an indirect way. Two situations are considered: a constant  electric field and an alternating electric field.

Let us consider an inertial coordinate
 system defined by the vector field $Z=\frac{\partial}{\partial t}$ corresponding to the laboratory frame.
The interaction of an ultra-relativistic bunch of particles with an external longitudinal
 electric field $E=(0,E_2(x),0)$ and zero magnetic field.
For narrow distributions it holds that
$\frac{dX^j}{dt}\frac{d\xi_j}{dt}=\mathcal{O}^1.$
This relation can be seen as follows.
For the linear dynamics $\xi=(\xi,0,-\xi,0)$ in the laboratory frame.
Also note that in the ultra-relativistic limit $\frac{dX^k}{ds}=(1+E,0,E,0)$,
with $E>> 1$.
Then the averaged Jacobi equation for the
limit $\epsilon^j \to  0$ in the ultra-relativistic regime correspond to the differential equations
\begin{align*}
0 & =\frac{d^2 \xi^i}{dt^2}+
2\frac{d\xi^j}{dt}\frac{dX^k}{dt}\Big(\frac{1}{2}({\bf F}^i\,_{j}
\frac{dX^m}{dt}\eta_{mk}+ {\bf
F}^i\,_{k}\frac{dX^m}{dt}\eta_{mj})\Big)\\
& +2\xi^l
\partial_l \Big(\frac{1}{2}({\bf F}^i\,_{j}\frac{dX^m}{dt}\eta_{mk}+ {\bf
F}^i\,_{k}\frac{dX^m}{dt}\eta_{mj})\Big)\\
& \cdot\frac{dX^j}{dt}\frac{dX^k}{dt}+
\big(\,^{\eta}\Gamma^i\,_{jk}+\xi^l\partial_l
\,^{\eta}\Gamma^i\,_{jk}\big)
\big(\frac{dX^j}{dt}\frac{dX^k}{dt}+2
\frac{dX^j}{dt}\frac{d\xi^k}{dt}\big).
\end{align*}
For the above longitudinal electric field, the equations of motion are
\begin{align*}
& 0 =\frac{d^2 \xi^0}{dt^2}+
\frac{d\xi^2}{dt}\frac{dX^k}{dt}E_2\langle y^m\rangle  \eta_{mk}+
\frac{d\xi^k}{dt}\frac{dX^2}{dt}E_2\langle y^m\rangle  \eta_{mk}\\
& +\xi^l \partial_l
\Big(\frac{dX^k}{dt}\frac{dX^2}{dt}E_2\langle y^m\rangle  \eta_{mk}+
\frac{dX^2}{dt}\frac{dX^j}{dt}E_2\langle y^m\rangle  \eta_{mj}\Big)\\
& +\big(\,^{\eta}\Gamma^0_{jk}+\xi^l\partial_l
\,^{\eta}\Gamma^0\,_{jk}\big)
\big(\frac{dX^j}{dt}\frac{dX^k}{dt}+2\frac{dX^j}{dt}
\frac{d\xi^k}{dt}\big),\\
& 0 =\frac{d^2 \xi^2}{dt^2}+
\frac{d\xi^0}{dt}\frac{dX^k}{dt}E_2\langle y^m\rangle  \eta_{mk}-
\frac{d\xi^k}{dt}\frac{dX^0}{dt}E_2\langle y^m\rangle  \eta_{mk}\\
& -\xi^l \partial_l
\Big(\frac{dX^j}{dt}\frac{dX^0}{dt}E_2\langle y^m\rangle  \eta_{mj}+
\frac{dX^0}{dt}\frac{dX^j}{dt}E_2\langle y^m\rangle  \eta_{mj} \Big)\\
&+\big(\,^{\eta}\Gamma^2\,_{jk}+\xi^l\partial_l
\,^{\eta}\Gamma^2\,_{jk}\big)
\big(\frac{dX^j}{dt}\frac{dX^k}{dt}+2\frac{dX^j}{dt}
\frac{d\xi^k}{dt}\big),\\
& 0 =\frac{d^2 X^1}{dt^2}+\big(\,^{\eta}\Gamma^1\,_{jk}+\xi^l\partial_l
\,^{\eta}\Gamma^1\,_{jk}\big)
\big(\frac{dX^j}{dt}\frac{dX^k}{dt}+
2\frac{dX^j}{dt}\frac{d\xi^k}{dt}\big),\\
& 0=\frac{d^2 X^2}{d t^2}+\big(\,^{\eta}\Gamma^2\,_{jk}+\xi^l\partial_l
\,^{\eta}\Gamma^2_{jk}\big)
\big(\frac{dX^j}{dt}\frac{dX^k}{dt}+2\frac{dX^j}{dt}
\frac{d\xi^k}{dt}\big).
\end{align*}
In an inertial coordinate system the inertial terms are zero. Therefore, the
system of equations in the linear longitudinal dynamics in the
ultra-relativistic regime simplifies to
\begin{align*}
& 0 =\frac{d^2 \xi^0}{dt^2}+
\frac{d\xi^2}{dt}\frac{dX^k}{dt}E_2\langle y^m\rangle  \eta_{mk}+
\frac{d\xi^k}{dt}\frac{dX^2}{dt}E_2\langle y^m\rangle  \eta_{mk}  +2\xi^l
\partial_l (\frac{dX^k}{dt}\frac{dX^2}{dt}E_2\langle y^m\rangle  \eta_{mk}),\\
& 0=\frac{d^2 \xi^2}{dt^2}-
\frac{d\xi^2}{dt}\frac{dX^k}{dt}E_2\langle y^m\rangle  \eta_{mk}-
\frac{d\xi^k}{dt}\frac{dX^0}{dt}E_2\langle y^m\rangle  \eta_{mk} -2\xi^l
\partial_l (\frac{dX^j}{dt}\frac{dX^0}{dt}E_2\langle y^m\rangle  \eta_{mj})\\
& 0= \frac{d^2 X^1}{dt^2},\quad 0=\frac{d^2 X^2}{d t^2}.
\end{align*}
If $\epsilon^k=\langle y^k\rangle  -\frac{dX^k}{dt}\approx 0$ and since
the distribution function has support on the unit hyperboloid,  $\langle y^k\rangle  \frac{dX_k}{dt}\approx 1+\alpha$.
Using also
the decoupling condition $\frac{dX^k}{dt}\frac{d\xi_k}{dt}\approx
0$, the first two equations are
\begin{equation}
\frac{d^2 \xi^0}{dt^2}+ \frac{d\xi^2}{ds}E_2 +2\xi^l
\partial_l (\frac{dX^2}{dt}E_2)=0,
\end{equation}
\begin{equation}
\frac{d^2 \xi^2}{dt^2}- \frac{d\xi^0}{dt}E_2-2\xi^l
\partial_l (\frac{dX^0}{dt}E_2)=0,
\label{equaciondinamicalongitudinal}
\end{equation}
Let us concentrate in equation (\ref{equaciondinamicalongitudinal}).
In the ultra-relativistic limit the velocity field
$\frac{dX^0}{dt}=\gamma(t) $ (in units where the speed of light is equal to $1$). Therefore, the equation above can be written as
\begin{equation}
\frac{d^2 \xi^2}{dt^2} +\frac{d\xi^2}{dt}E_2-2\gamma(t) \xi^l
\partial_l
E_2({X}+\xi^2)=0.
\label{longitudinalequation}
\end{equation}
Let us approximate by Taylor's expansion $E_2({X}+\xi^2)=E_2(X)+\xi^k\frac{\partial}{\partial \xi^k}E_2$ in equation (\ref{longitudinalequation}).
Due to the translational invariance of the partial derivatives,
$\partial_l \equiv\frac{\partial}{\partial \xi^l}$ in the above expressions, by the chain rule. Then we have
\begin{displaymath}
\frac{d^2 \xi^2}{dt^2} +\frac{d\xi^2}{dt}E_2-2\gamma(t)
\xi^k\frac{\partial}{\partial \xi^k}(E_2(X+\xi)-E_2(X))=0.
\end{displaymath}
If $E_2(X+\xi)$ can be
approximated linearly on $\xi$, using Euler's theorem of
homogeneous functions one gets the following expression:
\begin{align}
\frac{d^2 \xi^2}{dt^2} +\frac{d\xi^2}{dt}E_2(t)-2\gamma(t)
(E_2((X+\xi)-E_2(X))=0.
\label{longitudinaldynamics}
\end{align}
This is the Jacobi version of the linear longitudinal dynamics.
\subsection{Examples}
\begin{enumerate}

\item {\bf Constant longitudinal electric field.}

In this case the
equation of motion is
\begin{displaymath}
\frac{d^2 \xi^2}{dt^2} +\frac{d\xi^2}{dt}E_2=0.
\end{displaymath}
A particular solution is
\begin{displaymath}
\xi^2=-\frac{\xi^2}{E_2}(e^{-E_2(t-t_0)}-1).
\end{displaymath}
This equation is a linearized version of the longitudinal
dynamics \cite[Vol.I, chapter 8]{W}.

\item {\bf Alternate longitudinal electric field}. In this case,
the electric field is of the form
\begin{displaymath}
E_2(X^2+\xi^2)=E_2(0)sin(w_{rf}(X^2+\xi^2)).
\end{displaymath}
The differential equation is
\begin{displaymath}
\frac{d^2 \xi^2}{dt^2}
+\frac{d\xi^2}{dt}E_2(0)sin(w_{rf}(X^2+\xi^2))-2\gamma
E_2(0)(sin(w_{rf}(X^2+\xi^2))-sin(w_{rf}X^2))=0.
\end{displaymath}
We can expand this equation in $\xi$, since $\xi$ is small
\begin{displaymath}
\frac{d^2 \xi^2}{dt^2}
+\frac{d\xi^2}{ds}E_2(0)(sin(w_{rf}X^2)+cos(w_{rf}X^2)\xi^2)-2\gamma(t)\,
 E_2(0)(cos(w_{rf}X^2)\xi^2)=0.
\end{displaymath}
At first order in $\xi^2$ we have the equivalent expression
\begin{displaymath}
\frac{d^2 \xi^2}{dt^2}
+\frac{d\xi^2}{dt}E_2(0)sin(w_{rf}X^2)-2\gamma(t)\,
E_2(0)(cos(w_{rf}X^2)\xi^2)=0.
\end{displaymath}
We choose the initial phase such that $sin(w_{rf}X^2)\simeq 0$; therefore
$cos(w_{rf}X^2)\simeq 1$ and the equation is

\begin{equation}
\frac{d^2 \xi^2}{dt^2} -2\gamma(t)\,  E_2(0)\xi^2=0.
\end{equation}

\end{enumerate}

\section{Discussion}
We have shown that the
linear transversal beam dynamics in accelerator physics is obtained from the Jacobi equation of the averaged Lorentz connection
for electromagnetic fields ${\bf F(X+\xi)}$ linear in the deviation variables $\xi$. In particular we have proved that in the case when the magnetic fields are linear on $\xi$, like in a dipole and quadrupole magnetic fields, the transverse dynamics can be interpreted as the dynamics of the Jacobi equation of an affine, symmetric connection $\langle\,^L\nabla\rangle$. A similar conclusion follows for the linear longitudinal beam dynamics, where the effect of constant and oscillating electric fields on the bunch of particles are described as an approximation of the Jacobi equation for the averaged Lorentz connection $\langle\,^L\nabla\rangle$.

One advantage of the theory presented in this work respect to the standard treatment of beam dynamics based on the one particle Lorentz force equation is that our theory involves only
notions that are {observable}. In particular, the notion of reference trajectory that we provide is linked with the mean velocity field of a kinetic model, which is an observable quantity. We also describe a relation between the distribution
function and the collective behavior of the system.

The function $\langle Off^{1,3}_{\xi}\rangle  (t)$ measures the departure of the motion associated with the mean velocity vector field from the ideal reference trajectory associated with the trajectory of a point charged particle. 

Further theoretical research can be directed to generalize the approach to non-linear dynamics. This is apparently done by the so called {\it generalized Jacobi equations} for {\it affine} connections \cite{Perlick07}. Another direction is the study of conjugate points associated with the averaged Lorentz connection. Conjugate points can be associated with self-intersections of the solutions of the equation of motion. Finally, numerical implementation of the averaged Lorentz equation for numerical simulations can be interesting for accelerator particle physics, since its structure is more simple than the original Lorentz equation and it is expected better computational perspectives.

\paragraph{}
{\bf Acknowledgements.} Financially supported by EPSRC and Cockcroft Institute at the first version of this research at Lancaster University and by FAPESP process 2010/11934-6 at S\~ao Paulo University.

\small{
}

\end{document}